# *1/f* noise: a pedagogical review.

*Edoardo Milotti*

*Dipartimento di Fisica, Università di Udine and I.N.F.N. – Sezione di Trieste*
*Via delle Scienze, 208 – I-33100 Udine, Italy*

## 1. Introduction.

The triode was invented by Lee de Forest in 1907, and soon afterwards the first amplifiers were built. By 1921 the "thermionic tube" amplifiers were so developed that C. A. Hartmann [1] made the first courageous experiment to verify Schottky's formula for the shot noise spectral density [2]. Hartmann's attempt failed, and it was finally J. B. Johnson who successfully measured the predicted white noise spectrum [3]. However Johnson also measured an unexpected "flicker noise" at low frequency: his results are shown in figure 1, and shortly thereafter W. Schottky tried to provide a theoretical explanation [4].

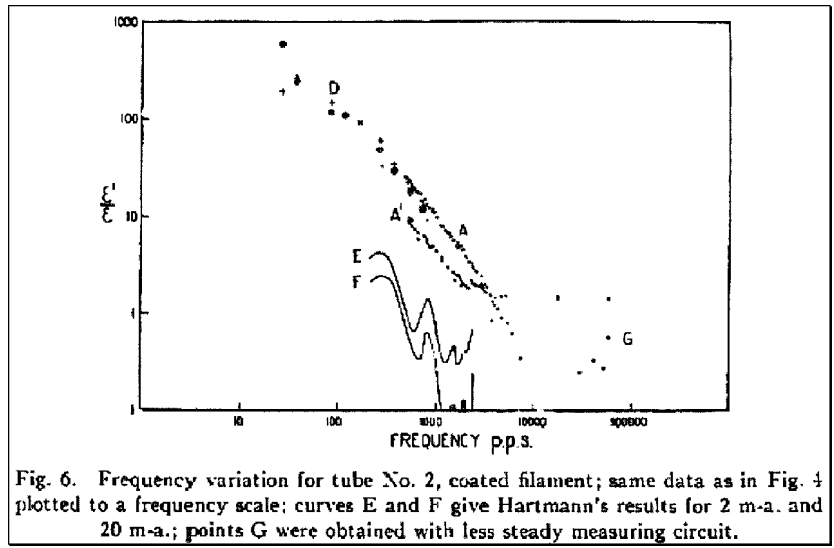

Figure 1: the spectral density observed by J. B. Johnson, as shown in his original paper [3]. The vertical scale represents the observed noise power density divided by the theoretical shot noise power density; the horizontal scale is the frequency in Hz.

Schottky's explanation was based on the physics of electron transport inside the vacuum tube, but in the years that followed Johnson's discovery of flicker noise it was found that this strange noise appeared again and again in many different electrical devices.
The observed spectral density of flicker noise is actually quite variable: it behaves like $1/f^{\alpha}$, where $\alpha$ is in the range $0.5 \div 1.5$, and usually this behavior extends over several frequency decades.
The appearance of power laws in the theory of critical phenomena and above all the work of B. Mandelbrot on fractals in the 1970's [5], seemed to indicate that something deeper was hidden in those ubiquitous spectra. Power laws and *1/f* spectra were found most unexpectedly in many different phenomena, and figure 2 shows two such spectra reproduced in a famous review paper by W. H. Press [6].

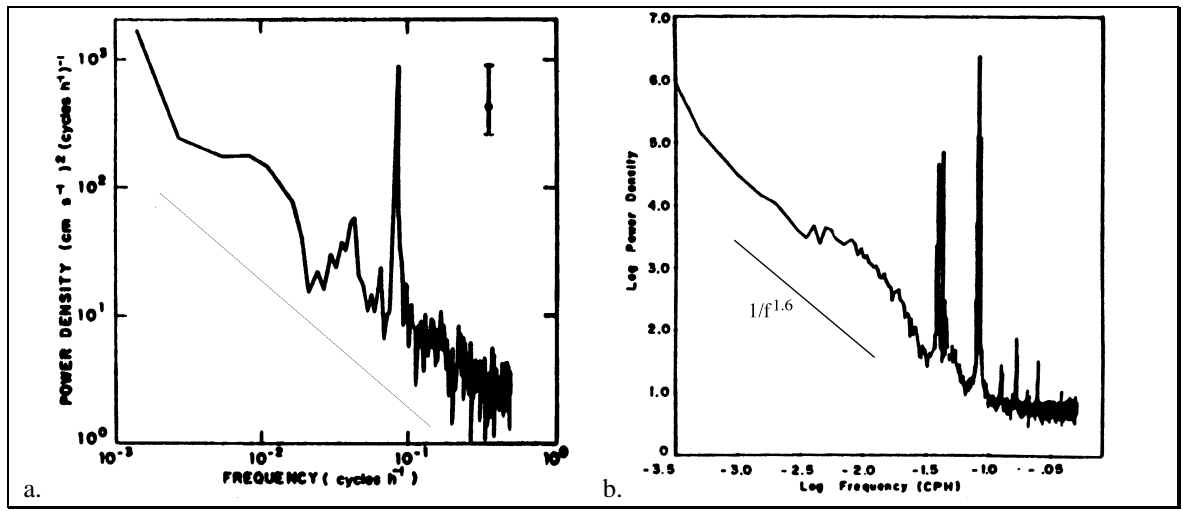

Figure 2: a. power spectrum of the east-west component of ocean current velocity [7]; the straight line shows the slope of a *1/f* spectrum. b. sea level at Bermuda: this is *1/f^α* spectrum with α ≈ 1.6 [8].

The work of Clarke and Voss on *1/f* noise in resistors also spawned an interesting aside, a study of *1/f* noise in music, which become widely known thanks to an excellent popularization made by M. Gardner in his *Scientific American* column [9]. Clarke and Voss found that both voice and music broadcasts have *1/f* spectra (see figure 3) [10], and they even devised an algorithm to compose "fractal" music [11].

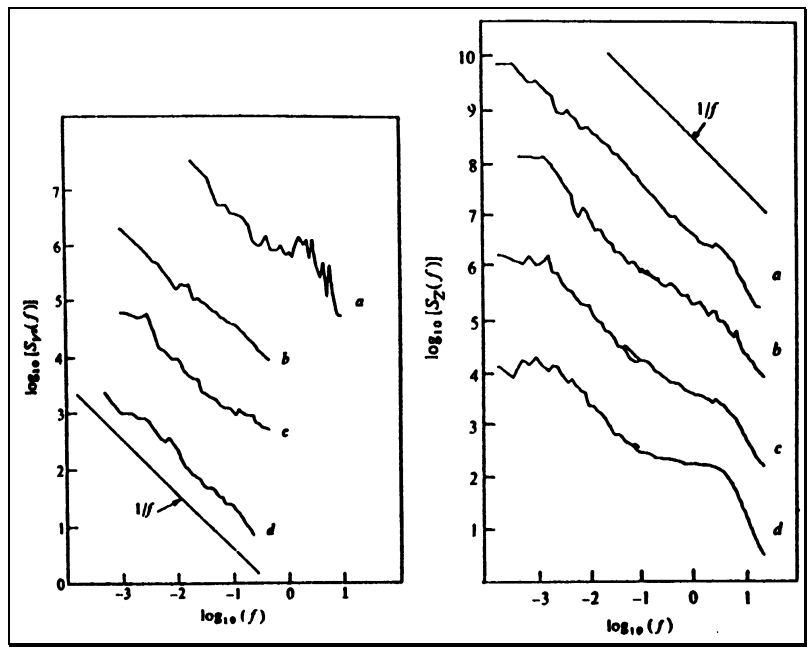

Figure 3: Loudness (left) and pitch (right) fluctuation spectra vs. frequency (Hz) (log-log scale), for a. Scott Joplin piano rags; b. classical radio station; c. rock station; d. news-and-talk station. (from ref. [10]).

By then many physicists were convinced that there had to be a deep reason for the ubiquity of this kind of power-law noises, that there had to be something akin to the universality of exponents in critical phenomena, and therefore many people set out to find an all-encompassing explanation.

## 2. $1/f^\alpha$ noise from the superposition of relaxation processes

An early and simple explanation of the appearance of $1/f^\alpha$ noise in vacuum tubes was implicit in some comments of Johnson [3], and was stated mathematically by Schottky [4]: there is a contribution to the vacuum tube current from cathode surface trapping sites, which release the electrons according to a simple exponential relaxation law $N(t) = N_0 e^{-\lambda t}$ for $t \geq 0$ and $N(t) = 0$ for $t < 0$. The Fourier transform of a single exponential relaxation process is

$$F(\omega) = \int_{-\infty}^{+\infty} N(t) e^{-i\omega t} dt = N_0 \int_0^{+\infty} e^{-(\lambda + i\omega)t} dt = \frac{N_0}{\lambda + i\omega} \tag{1}$$

therefore for a train of such pulses $N(t,t_k) = N_0 e^{-\lambda(t-t_k)}$ for $t \geq t_k$ and $N(t,t_k) = 0$ for $t < t_k$, we find

$$F(\omega) = \int_{-\infty}^{+\infty} \sum_k N(t,t_k) e^{-i\omega t} dt = N_0 \sum_k e^{i\omega t_k} \int_0^{+\infty} e^{-(\lambda + i\omega)t} dt = \frac{N_0}{\lambda + i\omega} \sum_k e^{i\omega t_k} \tag{2}$$

and the spectrum is

$$S(\omega) = \lim_T \frac{1}{T} \left\langle |F(\omega)|^2 \right\rangle = \frac{N_0^2}{\lambda^2 + \omega^2} \lim_T \frac{1}{T} \left\langle \left| \sum_k e^{i\omega t_k} \right|^2 \right\rangle = \frac{N_0^2 n}{\lambda^2 + \omega^2} \tag{3}$$

where $n$ is the average pulse rate and the triangle brackets denote an ensemble average. This spectrum is nearly flat near the origin, and after a transition region it becomes proportional to $1/\omega^2$ at high frequency. This was sufficient for Schottky, who had found such a dependence in Johnson's data, but later it became clear that a single relaxation process was not enough, and that there had to be a superposition of such processes, with a distribution of relaxation rates $\lambda$ [12]. If the relaxation rate is uniformly distributed between two values $\lambda_1$ and $\lambda_2$, and the amplitude of each pulse remains constant, we find the spectrum

$$S(\omega) = \frac{1}{\lambda_2 - \lambda_1} \int_{\lambda_1}^{\lambda_2} \frac{N_0^2 n}{\lambda^2 + \omega^2} d\lambda = \frac{N_0^2 n}{\omega(\lambda_2 - \lambda_1)} \left( \arctan \frac{\lambda_2}{\omega} - \arctan \frac{\lambda_1}{\omega} \right)$$

$$\approx \begin{cases} N_0^2 n & 0 < \omega \ll \lambda_1 \ll \lambda_2 \\ \dfrac{N_0^2 n \pi}{2\omega(\lambda_2 - \lambda_1)} & \lambda_1 \ll \omega \ll \lambda_2 \\ \dfrac{N_0^2 n}{\omega^2} & \lambda_1 \ll \lambda_2 \ll \omega \end{cases} \tag{4}$$

(see figures 4 and 5).

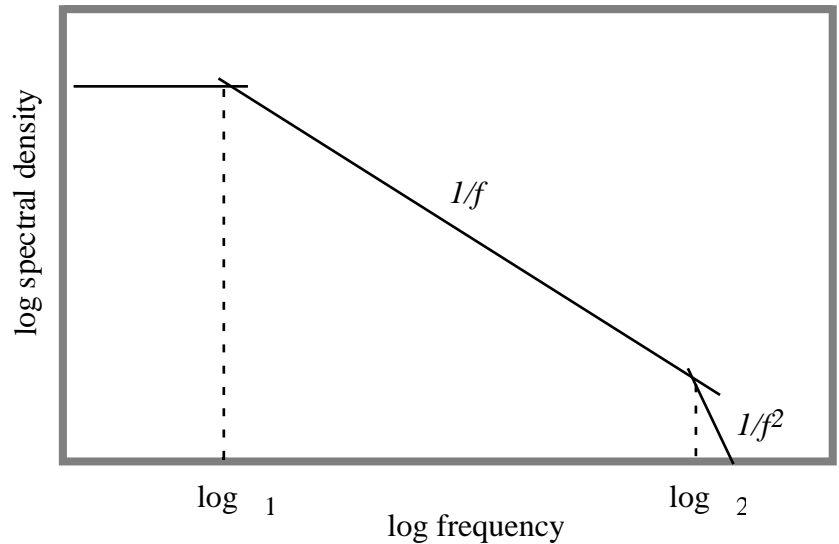

Figure 4: schematic shape of the spectral density (4). There are three characteristic regions: a white noise region at very low frequency, a *1/f* noise intermediate region and a *1/f²* region at high frequency.

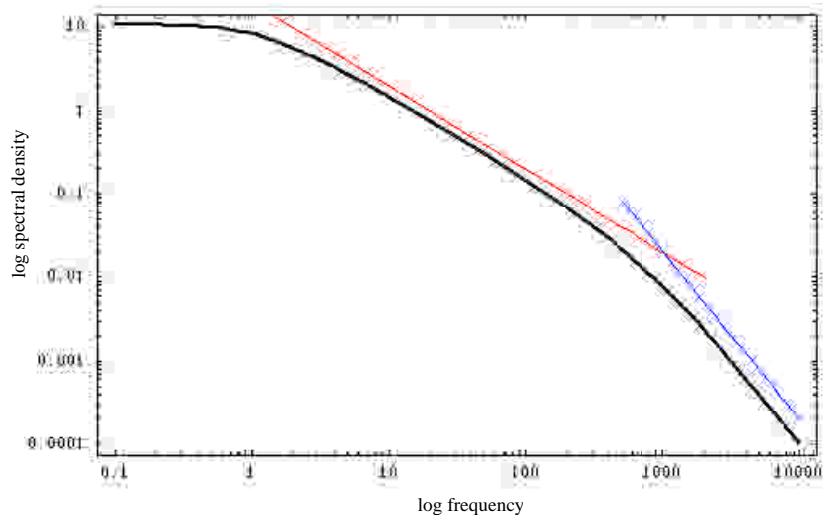

Figure 5: spectral density (arbitrary units) obtained from the superposition of 10000 relaxation processes with decay rates uniformly distributed and equally spaced between 1 and 1000 (arbitrary frequency units). The straight lines represent a *1/f* (red) and a *1/f²* (blue) spectral density.

How stable is this result? Is it still possible to obtain similar spectra changing the assumptions to fit reasonable physical needs?
These questions can at least partly be answered with a direct numerical simulation: figure 6 shows the result for a uniform random distribution of the relaxation rates, and it is clear that the final distribution is insensitive to small deviations from a perfectly uniform distribution of the relaxation rates.

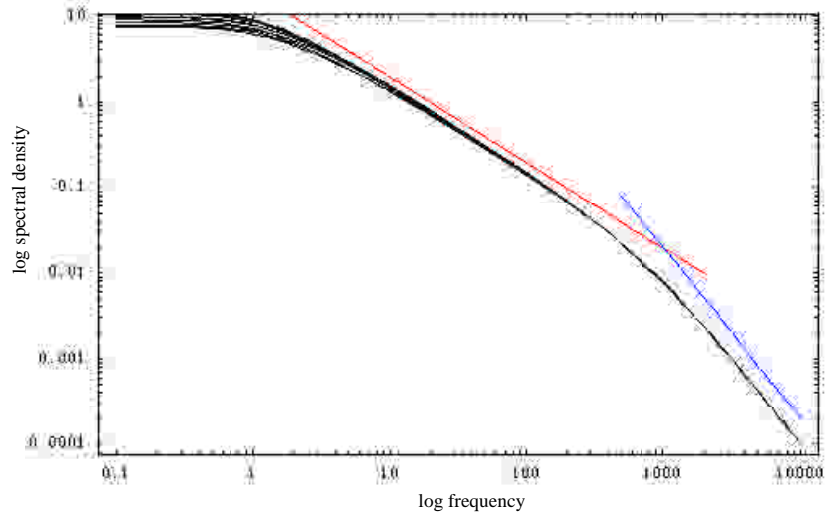

Figure 6: spectral densities (arbitrary units) obtained from the superposition of 10000 relaxation processes with decay rates randomly and uniformly distributed between 1 and 1000 (arbitrary frequency units). 10 spectral densities were generated and superposed on the same plot (black solid lines). The straight lines represent a *1/f* (red) and a *1/f²* (blue) spectral density.

We shall see later that the relaxation rates may be distributed according to different distributions, for instance we may have

$$dP(\lambda) = \frac{A}{\lambda^\beta} d\lambda \qquad (5)$$

in the range $\lambda_1 < \lambda < \lambda_2$: in this case it is still possible to integrate the spectrum exactly[1] and we obtain

$$S(\omega) \int_{\lambda_1}^{\lambda_2} \frac{1}{\lambda^2 + \omega^2} \frac{d\lambda}{\lambda^\beta} = \begin{cases} \frac{1}{\omega^2} \ln \frac{\lambda}{\sqrt{\lambda^2 + \omega^2}} \Big|_{\lambda_1}^{\lambda_2} & \text{if } \beta = 1 \\ \frac{\lambda^{1-\beta}}{(1-\beta)\omega^2} F\left(\frac{1-\beta}{2}, 1; 1+\frac{1-\beta}{2}; -\frac{\lambda^2}{\omega^2}\right) \Big|_{\lambda_1}^{\lambda_2} & \text{if } \beta \neq 1 \end{cases} \qquad (6)$$

where

$$F(a,b;c;z) = {}_2F_1(a,b;c;z) = \sum_{k=0}^{\infty} \frac{(a)_k (b)_k}{(c)_k} \frac{z^k}{k!} = \frac{\Gamma(c)}{\Gamma(b)\Gamma(c-b)} \int_0^1 t^{b-1}(1-t)^{c-b-1}(1-tz)^{-a} dt \qquad (7)$$

---

[1] Similar calculations have been done by Butz [13] and are summarized in the review paper by van der Ziel [14]

is the usual hypergeometric function. However we do not have to use the exact expression (6) to find the behavior of the spectral density in the range $\lambda_1 << \lambda << \lambda_2$, since we can approximate the exact integral as follows:

$$S(\omega) \approx \int_{\lambda_1}^{\lambda_2} \frac{1}{\lambda^2 + \omega^2} \frac{d\lambda}{\lambda^\beta} = \frac{1}{\omega^{1+\beta}} \int_{\lambda_1}^{\lambda_2} \frac{1}{(1+\lambda^2/\omega^2)} \frac{d(\lambda/\omega)}{(\lambda/\omega)^\beta} = \frac{1}{\omega^{1+\beta}} \int_{\lambda_1/\omega}^{\lambda_2/\omega} \frac{1}{(1+x^2)} \frac{dx}{x^\beta}$$

$$\approx \frac{1}{\omega^{1+\beta}} \int_0^\infty \frac{1}{(1+x^2)} \frac{dx}{x^\beta} \propto \frac{1}{\omega^{1+\beta}}$$

(8)

and thus we obtain a whole class of flicker noises with different exponents.

## 3. Infinitely large fluctuations?

From the previous discussion one may argue that it is important to find experimentally the actual limiting values $\lambda_1$ and $\lambda_2$, in order to characterize the noise process. Unfortunately this is seldom possible, and in most cases it seems that the *1/f* behavior continues as far as one can see: consider for instance the beautiful data of Pellegrini, Saletti, Terreni and Prudenziati[15] shown in figure 7, the *1/f* behavior extends over more than 6 frequency decades and there seems to be still no noise power flattening at low frequency.

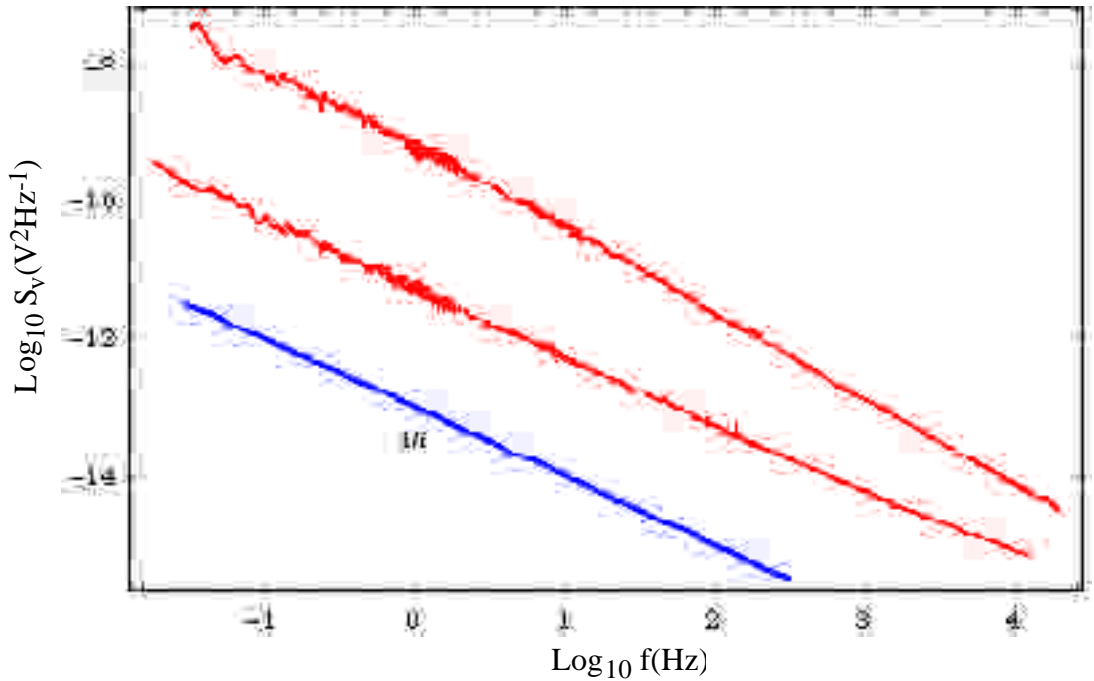

Figure 7. This figure shows the spectral densities of voltage fluctuations measured in two thin-film resistors by Pellegrini, Saletti, Terreni and Prudenziati [15]. They are very close to a perfect *1/f* noise, the behavior extends over more than 6 frequency decades and there seems to be no noise power flattening at low frequency.

Caloyannides did a very long data-taking run using operational amplifiers as noise sources and extended his measurements down to $10^{-6.3}$ Hz: he observed a $1/f^{1.23}$ spectrum with no flattening at low frequency (see figure 8) [16].

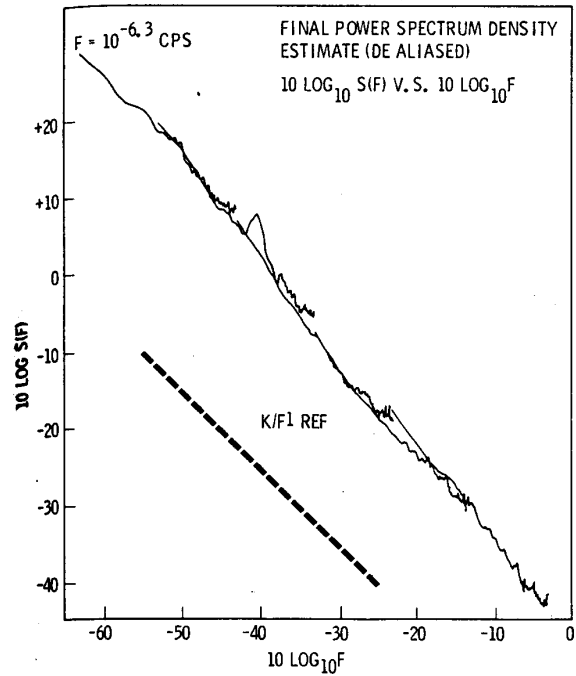

Figure 8: Caloyannides [16] took data for three months on an array of operational amplifiers, and measured a $1/f^{1.23}$ spectrum down to $10^{-6.3}$ Hz.

What if this behavior were real, and it continued indeed down to zero frequency? Then we would hit disaster, because the integrated fluctuation would be

$$\int_0^\infty S(f)df \sim \lim_{f_1 \to 0, f_2 \to \infty} \int_{f_1}^{f_2} \frac{1}{f^\alpha} df = \begin{cases} \lim_{f_1 \to 0, f_2 \to \infty} \ln \frac{f_2}{f_1} & \text{if } \alpha = 1 \\ \lim_{f_1 \to 0, f_2 \to \infty} \frac{f_2^{1-\alpha}}{1-\alpha} - \frac{f_1^{1-\alpha}}{1-\alpha} & \text{if } \alpha \neq 1 \end{cases} \quad (9)$$

and this expression always diverges, either at the low-frequency limit (for $\alpha>1$) or at the high-frequency limit (for $\alpha<1$) or both (for $\alpha=1$), so that low frequency fluctuations are arbitrarily large. Press commented jokingly on this divergence in the spectrum of the Bermuda sea level fluctuations (figure 2b) remarking that "If you postpone your Bermuda vacation for too long, the island may be underwater!" [6].

But is this divergence real? Flinn [17] produced a simple argument that shows that we should not worry about it even if it were there. Indeed for a true $1/f$ spectrum we know that

$$\int_{f_1}^{f_2} \frac{df}{f} = \ln \frac{f_2}{f_1} \quad (10)$$

so that the integrated fluctuation per decade is always the same. Moreover the lowest observable frequency is given by the inverse of the life of the Universe $\approx 2 \cdot 10^9$ years $\approx 6 \cdot 10^{16}$ s, and therefore it should be approximately $10^{-17}$ Hz. On the other hand it takes $\lambdabar_C/c \approx (4 \ 10^{-13} m)/(3 \ 10^{8} m/s) \approx 1.3 \ 10^{-21} s$ to go through an electron Compton wavelength at the speed of light, and this might be taken as the smallest observable time, which would correspond to a high frequency limit of $\approx 10^{21}$ Hz. There are 38 frequency decades between these two extremes, so that the highest possible total fluctuation can be only 38 times the total fluctuation between 1Hz and 10 Hz!

Even if we extend slightly Finn's argument, and take Planck's time as the smallest observable time $t_P = \frac{l_P}{c} = \sqrt{\frac{G\hbar}{c^5}} \approx 10^{-43} s$ we get a high frequency limit of $\approx 10^{43}$ Hz which yields a total of 59 frequency decades, the conclusion remains the same, and we should not worry too much about the mathematical divergences.

## 4. Noise in diffusion processes

Most of the early noise studies were carried out on resistors, operational amplifiers or other electronic equipment and systems. A special emphasis was placed on resistors, and thus it was quite natural to identify simple random processes like the simple random walk and more general diffusion processes as possible origins of *1/f* noise.

Let us start with the simple stochastic differential equation for the Wiener process

$$\frac{dx}{dt} = n(t) \qquad (11)$$

that describes Brownian motion in one space dimension, where *x(t)* is the position of the Brownian particle, and *n(t)* is a Gaussian white noise process with standard deviation σ, so that $\langle n(t)n(t+\tau)\rangle = \sigma^2 \delta(\tau)$. We recall that the spectral density of *n(t)* is just $\sigma^2/2\pi$, and since equation (11) implies the following relation between the Fourier transforms $X(\omega)$ and $N(\omega)$ of the two processes

$$-i\omega X(\omega) = N(\omega) \qquad (12)$$

we see that the spectral density of *x* is

$$S_x = \frac{\sigma^2}{2\pi\omega^2} \qquad (13)$$

and thus we conclude that Brownian motion has a *1/f²* spectrum.

A random walk in velocity space does not yield a very different result: now we start with the Langevin equation for the velocity *v(t)* with no external forces

$$\frac{dv}{dt} + \gamma v(t) = n(t) \qquad (14)$$

and then we find

$$-i\omega V(\omega) + \gamma V(\omega) = N(\omega) \qquad (15)$$

and the spectral density of the velocity

$$S_v = \frac{\sigma^2}{2\pi(\gamma^2 + \omega^2)} \tag{16}$$

The position $x(t)$ is the time integral of the velocity $v(t)$ and therefore the relationship between their respective spectra is

$$S_x = \frac{S_v}{\omega^2} \tag{17}$$

so that the spectrum of the position $x(t)$ is

$$S_x = \frac{\sigma^2}{2\pi(\gamma^2 + \omega^2)\omega^2} \tag{18}$$

which falls off like the spectrum of Brownian motion at low frequency, and much faster at high frequency. These formulas show that neither Brownian motion in position space nor Brownian motion in velocity space can adequately describe $1/f$ noise, at least in one space dimension only: the spectrum falls off too fast.

We know that Brownian motion describes a simple diffusion process, but maybe a more general diffusion process could do the trick and lead us to $1/f^\alpha$ spectra, so we drop the Langevin equation for Brownian motion and turn to the diffusion equation with a random driving term [19]:

$$\frac{\partial p(\mathbf{x},t)}{\partial t} = D \nabla^2 p(\mathbf{x},t) + f(\mathbf{x},t) \tag{19}$$

where $p$ is a probability density function and $f$ is a random driving term, and the observed process $r$ is related to $p$ by the integral formula

$$r(t) = \int_V g(\mathbf{x}) p(\mathbf{x},t) d^3 x \tag{20}$$

where $g$ is a coupling factor. Let $P$ and $F$ be the Fourier transforms defined by

$$p(\mathbf{x},t) = \frac{1}{2\pi} \int_{-\infty}^{+\infty} \int_V P(\mathbf{q},\omega) e^{-i(\omega t - \mathbf{q}\cdot\mathbf{x})} d^3 q d\omega$$
$$f(\mathbf{x},t) = \frac{1}{2\pi} \int_{-\infty}^{+\infty} \int_V F(\mathbf{q},\omega) e^{-i(\omega t - \mathbf{q}\cdot\mathbf{x})} d^3 q d\omega \tag{21}$$

then we obtain

$$P(\mathbf{q},\omega) = \frac{F(\mathbf{q},\omega)}{-i\omega + Dq^2} \tag{22}$$

and the spectral density at a given position $\mathbf{x}$ is

$$S_r(\omega) \propto \left| \int \frac{G^*(\mathbf{q}) F(\mathbf{q},\omega)}{-i\omega + Dq^2} d^3 q \right|^2 \tag{23}$$

where $G$ is the (space) Fourier transform of $g$. If the driving term is local, so that $\langle f(\mathbf{x},t)f(\mathbf{x}',t')\rangle = A(\mathbf{x},t,t')\delta(\mathbf{x}-\mathbf{x}')$, equation (23) simplifies to

$$S_r(\omega) \propto \int \frac{|G(\mathbf{q})|^2 A(\mathbf{x},t,t')}{\omega^2 + (Dq^2)^2} d^3q \tag{24}$$

It is not difficult to derive a *1/f* spectrum from expression (24) with the proper assumptions on $G$ and $A$, but it is much more difficult to give physical meaning to these mathematical assumptions (see the discussion in [19]). Several authors (including myself [20]) have tried to give meaning to the diffusion approach, but the final word has not been written yet.

Before moving on to the next topic it is important to notice that the diffusion equation can be easily discretized on a square lattice, and in 2 space dimensions we obtain the following discrete version:

$$p_{j,k}(t+\Delta t) - p_{j,k}(t) = D\frac{\Delta t}{(\Delta x)^2}\left(p_{j+1,k}(t) + p_{j-1,k}(t) + p_{j,k+1}(t) + p_{j,k-1}(t) - 4p_{j,k}(t)\right) + f_{j,k}(t) \tag{25}$$

We shall return to equation (25) in section 6.

## 5. Statistical properties

The noise spectrum is certainly important, but it is not all: already J. B. Johnson in his 1925 experiment [3] looked for other characteristics of flicker noise in vacuum tubes, like the dependence on total current of the noise density, etc.

And indeed, the spectral density characterizes a noise process completely only if the process is stationary, ergodic and Gaussian: does the observed *1/f* noise satisfy all these constraints? How else can we characterize *1/f* noise?

These questions are answered at least partly in the literature on *1/f* noise in electronic devices, and here I concentrate on three important properties[2].

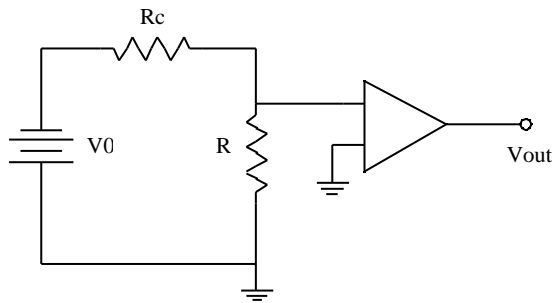

Figure 9: typical circuit used to measure voltage (or equivalently current or conductance) noise in the resistor R.

---

[2] Most of the early research was centered on conductors, and other properties like isotropy were tested: the interested reader can turn to the excellent reviews of Dutta and Horn [19], Weissmann [21] and Kogan [22].

### 5.1 Is noise present at equilibrium?

The work of many physicist and in particular of F. N. Hooge and collaborators [23], produced several empirical formulas for *1/f* noise in resistors, and in particular Hooge [24] showed that the *1/f* voltage spectral density can be parametrized by the formula

$$S_V(f) = \gamma \frac{V_{DC}^{2+\beta}}{N_c f^\alpha} \tag{26}$$

where $\alpha$, $\beta$ and $\gamma$ are constants, $V_{DC}$ is the applied voltage and $N_c$ is the total number of charge carriers in the sample. This formula relates *1/f* noise to the passage of current in the sample, and so people asked whether the noise was still present without a driving current. The problem was settled experimentally by Clarke and Voss [25] who found that *1/f* noise was indeed present at equilibrium and this result was later confirmed by Beck and Spruit [26]. The conclusion is true with a caveat: in the experiment there was no driving current, but there was also no guarantee of thermal equilibrium.

### 5.2 Is the noise process stationary?

Let us recall that a stochastic process $x$ is stationary in a strict sense when the ensemble averages $\langle x(t) \rangle$, $\langle x(t)x(t+\tau_1) \rangle$, $\langle x(t)x(t+\tau_1)x(t+\tau_2) \rangle$, ..., do not depend on time. We have already seen that *1/f* noise has some disturbing divergences: in practical terms this means that when we take an experimental record it may grow without limits, and it may superficially look like a deterministic process with the addition of white noise, i.e. *1/f* noise may easily simulate an instrumental drift with a time-varying average, so that one may wonder if this noise is actually stationary.

In 1969 Brophy [27] found that the zero-crossing statistics was Poissonian, just what one would expect from a stationary process, while shortly later Brophy and Greenstein [28] reported traces of nonstationarity. A few other papers followed, untili in 1975 Stoisiek and Wolf found that the statistical properties of *1/f* noise were "fully consistent with the assumption of stationarity".

### 5.3. Is the process Gaussian?

There are two important reasons to answer this question: in the first place a Gaussian process is completely characterized by its average value and by the spectral density; secondly linear processes – like simple diffusion – are always Gaussian, therefore Gaussianity is an important clue of linearity at the microscopic level.

To demostrate linearity Voss [29] produced experimental plots of the quantity $\langle V(t)|V_0\rangle / V_0$ in several conductors, and he related these plots to both linearity and Gaussianity: however these plots do not really demonstrate linearity, they just show that the noise processes observed by Voss were reasonably Gaussian..

Later several groups reported non-Gaussian signals (see, e.g., [30]), but Gaussianity can always be recovered at the macroscopic level from the superposition of many microscopic processes (this is a manifestation of the central limit theorem).

From the theoretical point of view linearity is not welcome, since many attempts to understand *1/f* noise are actually based on nonlinear processes.

Notice that the properties that we have discussed in this section have been tested on conductors only, i.e. on one very special subset of the systems that display *1/f* noise, therefore we cannot draw any general conclusion from these observations.

# 6. Self organized criticality

The outstanding feature of *1/f* noise is that it is scale invariant, i.e. it looks the same for any choice of frequency or time unit, and for this reason it has been widely considered a prominent manifestation of the fractal character of many natural phenomena. Since many nonlinear processes have complex phase spaces with fractal attractors, several physicists have looked into nonlinear processes as sources of *1/f* noise.

In 1987 Bak, Tang and Wiesenfeld (BTW) introduced a nonlinear model system that was met with wide interest and has since generated spates of scientific papers, the so-called *Sandpile Model* [31,32]: the title of the original paper was *Self Organized Criticality: An Explanation of 1/f Noise*, and it put forward a very ambitious program, as it described a nonlinear process that had fractal characteristics, a complex behaviour that mimicked a noise process, a spectral density that the authors claimed to be *1/f*, and displayed a limiting behavior that was called *self organized criticality*.

The sandpile model is closely related to an earlier classical model developed to describe the charge-density waves observed in some special conductors like $NbSe_3$ or $K_{0.3}MoO_3$ (see [33] and [34] for very readable reviews or [35] for a more technical one): in this model a charge-density wave may be viewed as a single particle (or an array of particles) in a periodic or quasi-periodic potential (see figure 10).

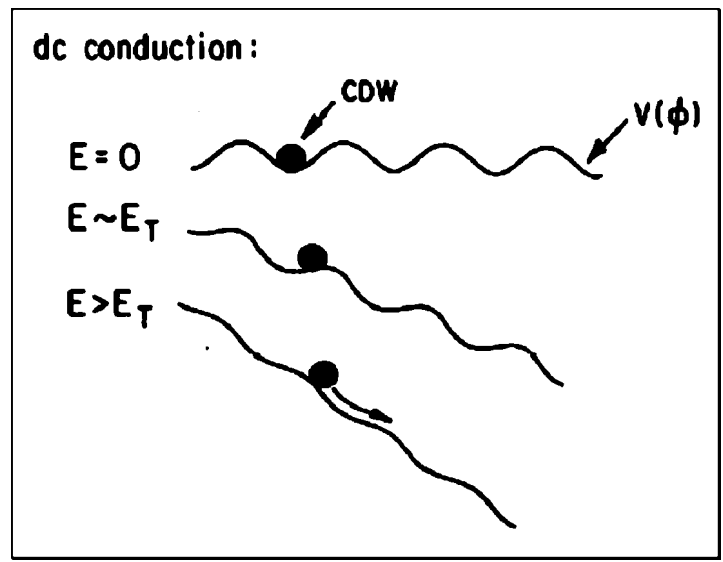

Figure 10: classical model for charge-density-wave conduction (from [35]): when pinning impurities are present the troughs have variable depths. Here a charge-density wave may be viewed as a single particle in a periodic or quasi-periodic potential and this simplified model closely resembles the "many-pendula" model introduced by Bak, Tang and Wiesenfeld in [31] to describe self-organized criticality.

BTW introduced a coupled-pendulum model of self organized criticality but the analogy with sand proved to be so suggestive, that soon only the sandpile paradigm survived in the literature[3]:. The two-dimensional version of the model is described mathematically by the discrete evolution equations

---

[3]Notice that a "sand model" of *1/f* noise had been described as early as 1974 by Schick and Verveen [36]

$$\begin{cases} z_{j,k} \to z_{j,k} & \text{if } z_{j,k} \le K \\ \begin{aligned} z_{j,k} &\to z_{j,k} - 4 \\ z_{j\pm1,k} &\to z_{j\pm1,k} + 1 \\ z_{j,k\pm1} &\to z_{j,k\pm1} + 1 \end{aligned} & \text{if } z_{j,k} > K \end{cases} \qquad (27)$$

where $z_{j,k}$ is an integer variable that may be taken to represent the height differences between adjacent nodes in a two-dimensional lattice, and $K$ is a threshold value. When we compare equations (25) and (27) we see that the dynamics described by (27) represent a discrete, nonlinear diffusion process.

Being discrete and nonlinear, the sandpile model is very hard to study with the usual analytical tools, but it can easily be adapted to large scale numerical calculations and in [32] BTW reported $1/f^{\alpha}$ spectral densities obtained from their numerical simulations, with $\alpha$ near 1.

BTW also argued that the sandpile model exhibits a form of self-organization as the slope of the sandpile approaches a limiting value, just as in real sandpiles (in plain words the diffusion process cannot proceed until the threshold $K$ is reached) and they dubbed it Self Organized Criticality (SOC).

Papers [31] and [32] initiated a flurry of activity as many physicists tried to replicate BTW's results, either with their own numerical simulations or with direct simulations of sand flows, and sometimes also with more sophisticated analytical tools like the renormalization group approach. However most of these studies have widened the scope of application of SOC and have not led to a deeper understanding of the behavior of the model, which remains controversial. For instance, there are applications of SOC to fields as diverse as fire propagation[37] or evolutionary biology[38,39], but alongside the enthusiastic attitudes of some physicists (see, e.g. [40]) there is an increasing number of papers that raise doubts on the model validity and actual applicability.

The study of $1/f$ noise started in electronics, and we have already seen that electrical $1/f$ noise seems to be stationary and Gaussian, but the pulses of sandpiles are not Gaussian, and the other statistical properties differ from those of the observed electrical $1/f$ noise as well [41]; moreover electrical $1/f$ noise exists at equilibrium while all SOC models require an external driving process. Another point that remains highly controversial is the reality of the claimed $1/f$ spectra: are they $1/f^{\alpha}$ spectra (with $\alpha \approx 1$) or are they trivial $1/f^2$ spectra? If they were just $1/f^2$ spectra, then SOC would not be essentially different from a simple Brownian process (see section 4), and therefore it is essential to establish the numerical value of the exponent. Jaeger, Liu and Nagel performed an experiment with a real sandpile in 1988 [42,43] and their results contradicted BTW's claims, they did not observe a $1/f^{\alpha}$ spectrum, but rather a white noise plateau – modified at low frequency by grain size effects – followed by a $1/f^2$ tail.

Jensen, Christensen and Fogedby added mathematical substance to the measurements of Jaeger, Liu and Nagel [43], and they demostrated that the spectra in real sandpiles are actually $1/f^2$ [44]. The problem of establishing the exponent has been aggravated by serious errors in the literature, like in ref. [45], where the authors claimed to have found a sandpile model with a $1/f$ spectrum, while they actually found a $1/f^2$ spectrum and forgot to square the Fourier coefficients.

The importance of finding $1/f^{\alpha}$ spectra with $\alpha \approx 2$ seems to have escaped the attention of some authors like Dalton and Corcoran, who recently performed another experiment on a granular system [46], found a trivial $1/f^2$ spectrum and conclude that this is in accordance with SOC.

On the whole it does not seem that SOC adequately describes *1/f* noise, nor that it can actually aspire to the universality that its supporters claim[4].

## 7. Earthquakes

After Mandelbrot's work, *1/f* noise has often been associated to fractal phenomena and other power laws, and the physics of earthquakes is just one of those fields where the concepts of scaling – which leads to power laws – and later of SOC, seem to be applicable. The mainstay of this field is the celebrated Gutenberg-Richter law [48] which states that *N(M)*, the number of earthquakes with magnitude greater than *M*, is proportional to $10^{-bM}$, i.e. $\log_{10} N(M) = A - bM$ where the slope *b* is found to be a number near 1. There is also a classical earthquake model with dissipative nonlinear dynamics, the Burridge-Knopoff model [49], so that the challenge lies in solving the Burridge-Knopoff model – or another similar model – to retrieve the Gutenberg-Richter law, and other earthquake statistics.

I have used the data available from an online earthquake database [50], to plot a Gutenberg-Richter distribution for the year 2000, and the result is shown in figure 11.

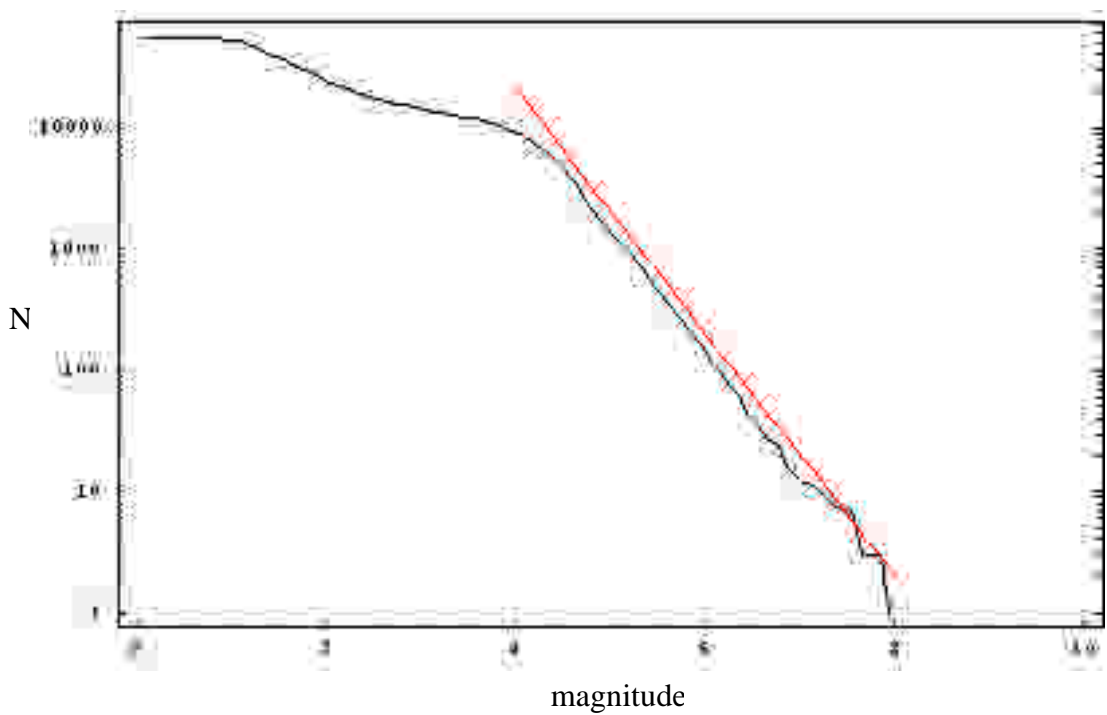

Figure 11: cumulative distribution *N* of the earthquakes worldwide for the year 2000 from the Berkeley database [50]: the red line has slope -1. Since the earhtquake magnitude is a logarithmic measure of the energy released by the earthquake, this may be viewed as a log-log plot. Notice that the power law region has a limited extension, approximately 3 decades, i.e. considerably less than the power law region in resistor noise spectra.

---

[4] One may also wonder if sandpiles are as universally applicable as the supporters claim: see ref. [47] for a historian's opinion.

Figure 11 shows that there is indeed a magnitude range where a power law holds, while there is a low magnitude region where the power law breaks down. Does the Burridge-Knopoff model explain all of this? The Burridge-Knopoff model is a simple block-and-spring model of a crustal fault (see figure 12) originally introduced in 1967, and it has been extensively studied [51], especially by Carlson and collaborators [52,53].

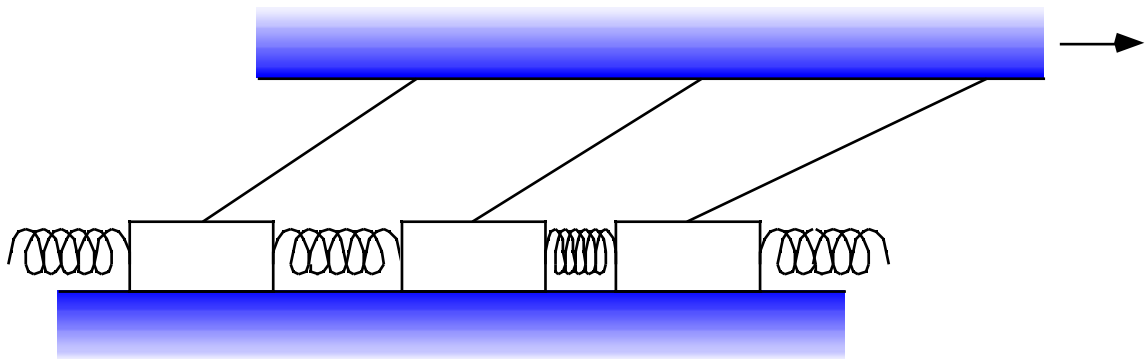

Figure 12: in the Burridge-Knopoff model there are blocks that slide on a plane with friction, as they are drawn forward by an upper plate that moves with constant speed. Each block is connected to the adjacent blocks by springs. This is the simplified model of a fault introduced by Burridge and Knopoff in 1967 [49]: a nice JAVA simulation is also available online [54].

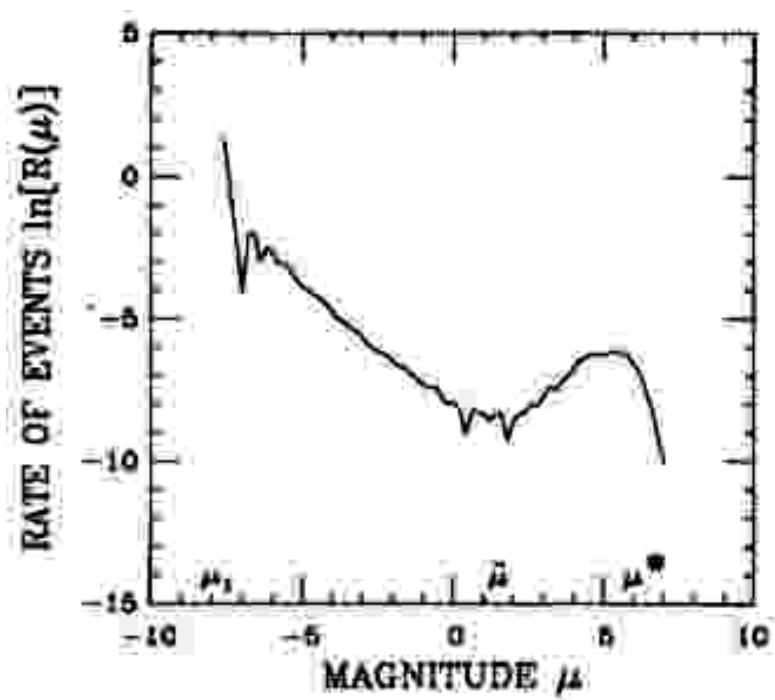

Figure 13: frequency distribution of the slip events (earthquakes) of magnitude µ taken from [53]. Notice the large bump that corresponds to an excess of events of high magnitude.

Figure 13 shows the results of the simulations in [53]: it looks like the intuitive Burridge-Knopoff model breaks down at high magnitude. Actually the plot in figure 13 is the frequency distribution of slip events, not the cumulative distribution as in the Gutenberg-Richter law, but the derivative of a power law is stilla a power law, so one would not expect a very different situation if one took the frequency distribution of earthquakes. However this is not so, because the Gutenberg-Richter law is a power law only in a rather limited magnitude range, and the frequency distribution of earthquakes – the same quakes that were used to produce figure 11 – has a very different look (figure 14), very similar to the one in figure 13.

Earthquakes are complex phenomena, and the restricted power law range may come from some fundamental difference in physics between small and large magnitude earthquakes [55,56]: here we can conclude that whatever the origin of the breakdown, the concept of scale invariance seems to have a very limited applicability to earthquakes.

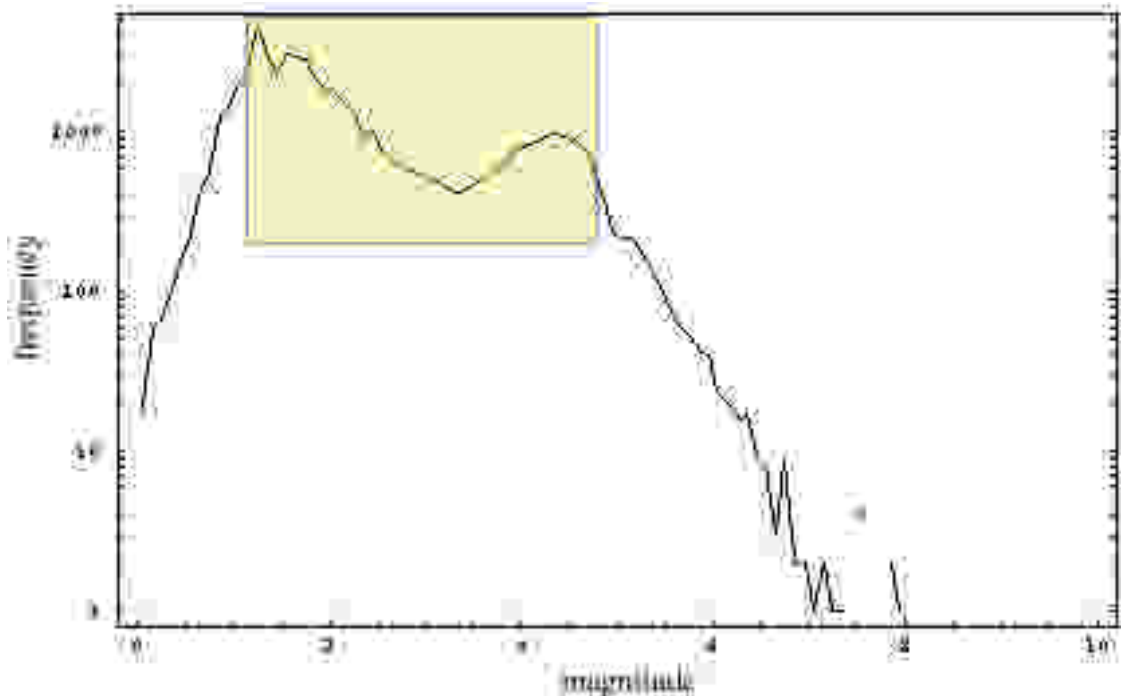

Figure 14: frequency distribution of the earthquakes worldwide for the year 2000 from the Berkeley database [50] (histogram bins have a 0.1 magnitude width): notice the large bump around magnitude 4. Should we identify the boxed region with the frequency distribution shown in figure 13?

## 8. *1/f* noise in deterministic dynamical systems

One explanation of the occurrence of *1/f* noise in resistors is that the charge carriers get trapped in capture sites and are released with variable rates (this is essentially the extension of Schottky's original explanation of *1/f* noise in vacuum tubes), and in the search for a "universal" explanation of *1/f* noise one may wonder if the same mechanism may act in more general dynamical systems. Geisel, Zacherl and Radons (GRZ) [57] devised just one such mechanism : it is well known that if we are given a random Hamiltonian its phase space splits in chaotic regions where the system point follows pseudoorbits and in ordered regions where the system point follows periodic orbits (this is the essence of the KAM theorem). The ordered regions are often surrounded by a hierarchi of cantori, and the conjecture of GRZ is that the system point gets temporarily trapped in these cantori and is released with variable rates, just as the charge carriers in an ordinary conductor with trapping sites. GRZ considered in particolar a classical particle in a periodic two-dimensional potential

$$V(x,y) = A + B(\cos x + \cos y) + C\cos x \cos y \tag{28}$$

(see figure 15), and solved numerically the coupled equations of motion

$$\begin{aligned}\ddot{x} &= (B + C\cos y)\sin x \\ \ddot{y} &= (B + C\cos x)\sin y\end{aligned} \tag{29}$$

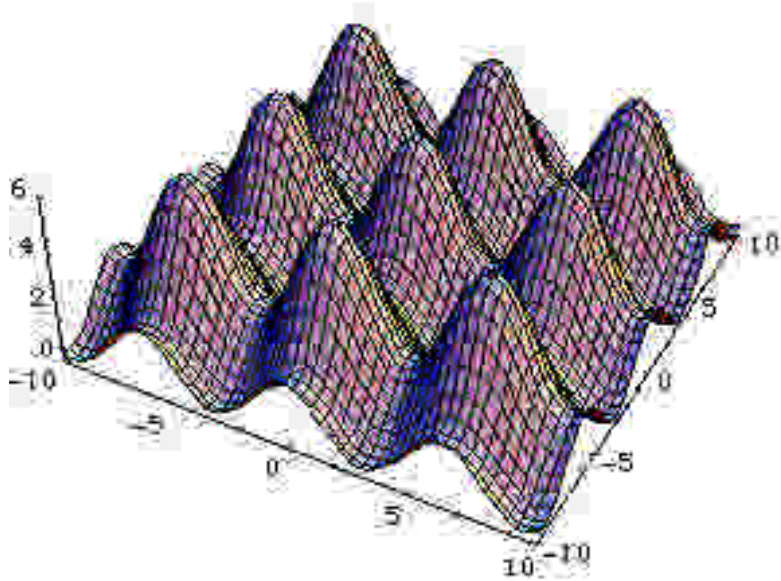

Figure 15: two-dimensional periodic potential used by Geisel, Zacherl and Radons [57]. GRZ took A=2.5, B=1.5 and C=0.5, and the total energy E=4, so that the maximum value of the potential is 6, and the particle performs a complex motion amid the peaks of the potential.

The particle performs a complex motion amid the many peaks of the potential, which looks like a sort of random walk (figure 16), while the velocity of the particle looks like a periodic signal plus noise: figure 17 shows the spectral density of the *y* velocity fluctuations which do have a peak plus a *1/f* noise background before the peak.

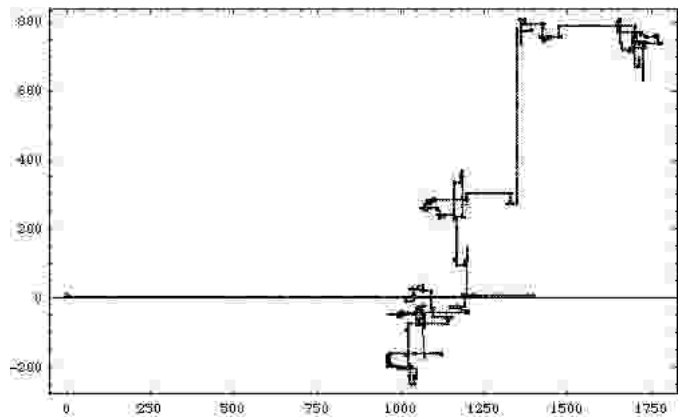

Figure 16: plot of the (*x*,*y*) position of the particle, which starts from (1.5,1.5) in this simulation: the particle performs a kind of random walk.

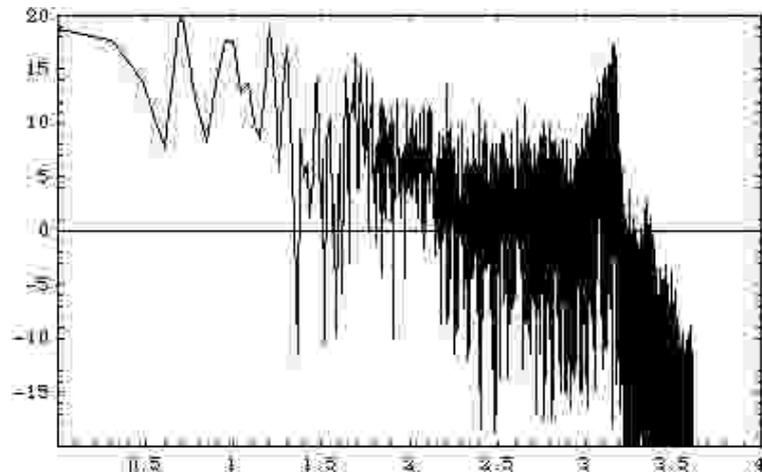
Figure 17: plot of the logarithm of the spectral density of the *y* velocity (dBrms) vs. log frequency (both scales in arbitrary units) for a numerical solution of the GRZ model Hamiltonian. There is a *1/f* part followed by a peak from a periodic signal (the system point orbits around limit cycles).

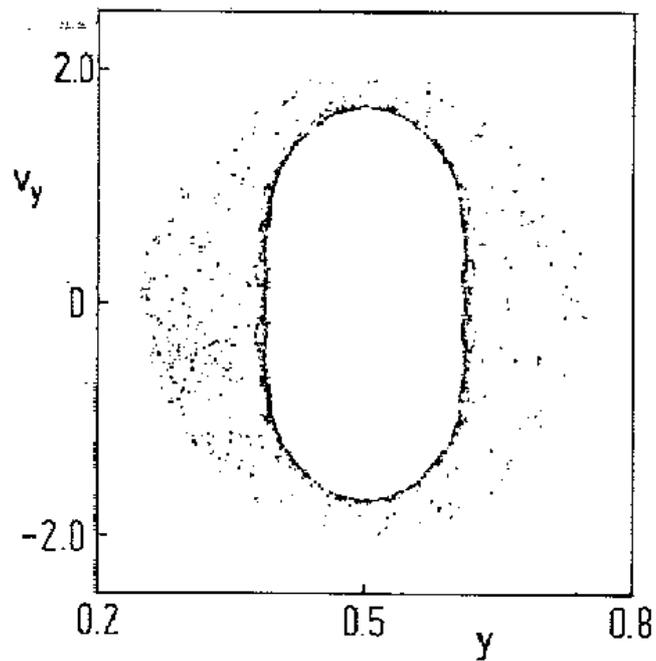
Figure 18: Poincaré surface of section for the GRZ model Hamiltonian (obtained for $x = 0 \mod 2\pi$), from ref. [57].

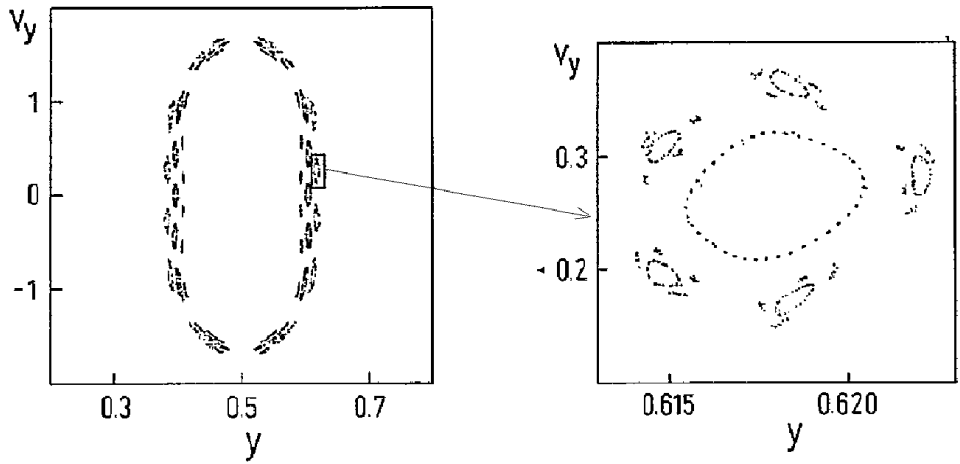

Figure 19: there is a whole hierarchi of cantori around the central island in figure 17. According to GRZ the particle is trapped in these cantori and this produces the observed *1/f* spectral density (adapted from ref. [57]).

The dynamical system of GRZ is a Hamiltonian system, i.e. it belongs to a specific subclass, however in physics there are non-Hamiltonian systems as well, like the celebrated Lorentz system

$$\dot{x} = \sigma(y - x)$$
$$\dot{y} = -xz + rx - y \quad\quad\quad (30)$$
$$\dot{z} = xy - bz$$

which has been extensively studied since its first appearance in 1963 [58,59]. Systems like this have many interesting features, such as stable limit cycles, strange attractors, intermittency and *1/f* noise. Intermittent behavior is characterized by a seemingly stationary signal interrrupted by bursts of activity, as in figure 20 (see refs. [60,61,62]).

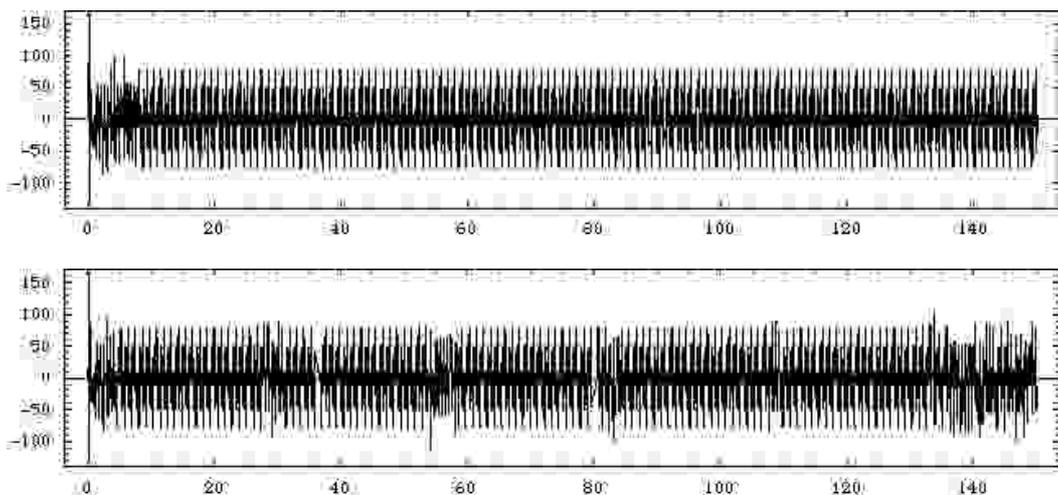

Figure 20: *y* amplitude vs. time (both arbitrary units) in the Lorentz model for $\sigma = 10$, $b = 8/3$ and $r = 166$ (upper plot), $r = 166.1$ (lower plot).

The fundamental work of Feingebaum on the transition to chaos in the logistic map has shown that simple functions may be used as representatives of a much wider class of functions [63]: the function

$$f(x) = x + ux^2 (\mathrm{mod}\, 1) \tag{31}$$

which maps the (0,1) interval into itself has been used to study both intermittence and power-law noise in dynamical systems [61,64,65]. Repeated iteration of the map (31) yields a sequence

$$x_{n+1} = f(x_n) = x_n + ux_n^2 (\mathrm{mod}\, 1) \tag{32}$$

which has both an intermittent behavior and a very clean *1/f* spectral density (figure 21). Schuster and Procaccia have [63] shown theoretically that this is a real *1/f* spectral density using the Feingebaum renormalization group method in function space [63].

It is important to notice that the fluctuations produced by the dynamical systems discussed in this section are far from Gaussian, and therefore the spectral densities are not sufficient to characterize the processes.

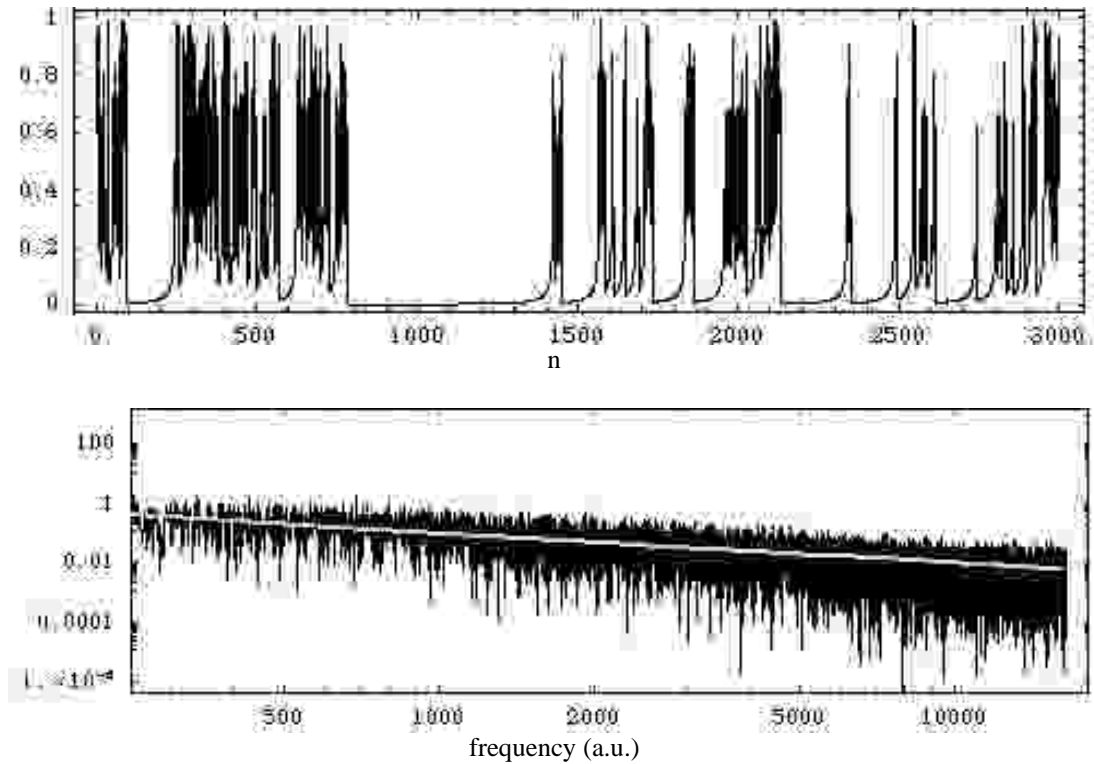

Figure 21: intermittent behavior in the iterated map defined by equations (31) and (32) (upper plot), and spectral density from a numerical calculation (lower plot); here *u*=1. The gray line shows the ideal *1/f* behavior.

## 9. An example from biophysics

As an example of a biophysical system in which *1/f* noise is actually observed, I wish to mention the fluctuations of the electrical dipole moment of lysozome, an important enzyme.

This is a very interesting physical system inasmuch as it provides information on the structure of water and on its interaction with biological molecules [66]. The fluctuations have been studied by Careri and Consolini [67] and the spectral density is once again a $1/f^\alpha$ spectrum, with $\alpha \approx 1.5$ (see figure )

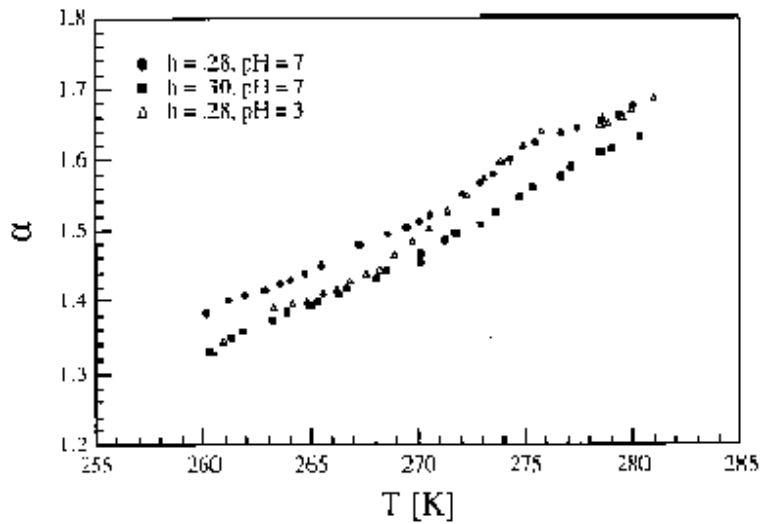

Figure 22: high-frequency noise exponent for the $1/f^\alpha$ spectral density of electric dipole fluctuations in hydrated lysozome powder as a function of temperature, as measured by Careri and Consolini [67].

We believe that the electrical dipole fluctuations are due to the migration of free protons on the molecule surface, and to test this model we have set up a Monte Carlo simulation [68]. Figure 23 shows a simulated dipole signal, and the average spectral density for many such signals: the first results obtained with this approach are encouraging, as we find good values for the noise exponent and reasonable dependencies on the other physical parameters. The spectral density in this model is a good example of the discussion of section 2, since it is found that the distribution of transition rates between different proton states follows a power law, and therefore produces a $1/f^\alpha$ spectral density. The individual signals are non-Gaussian, but the macroscopic signal is a superposition of many such signals, and thus it is Gaussian.

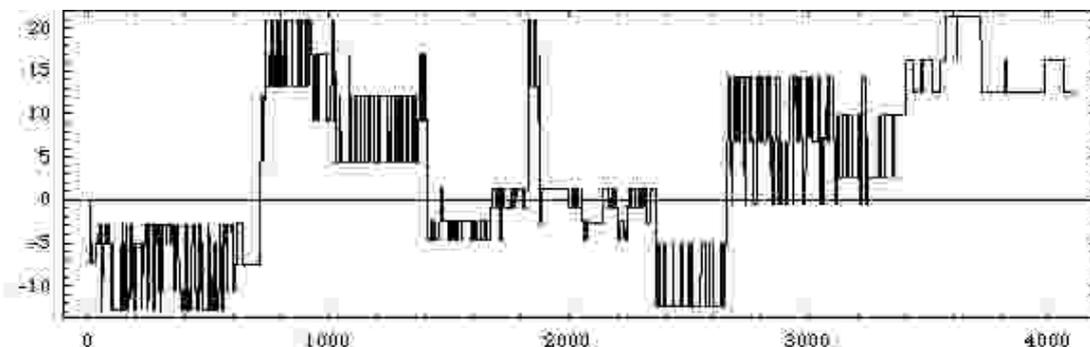

Figure 23: simulated dipole signal (*x* component) for a single molecule. Signals such as this are seen in conventional conductors as well (see, e.g. [69]).

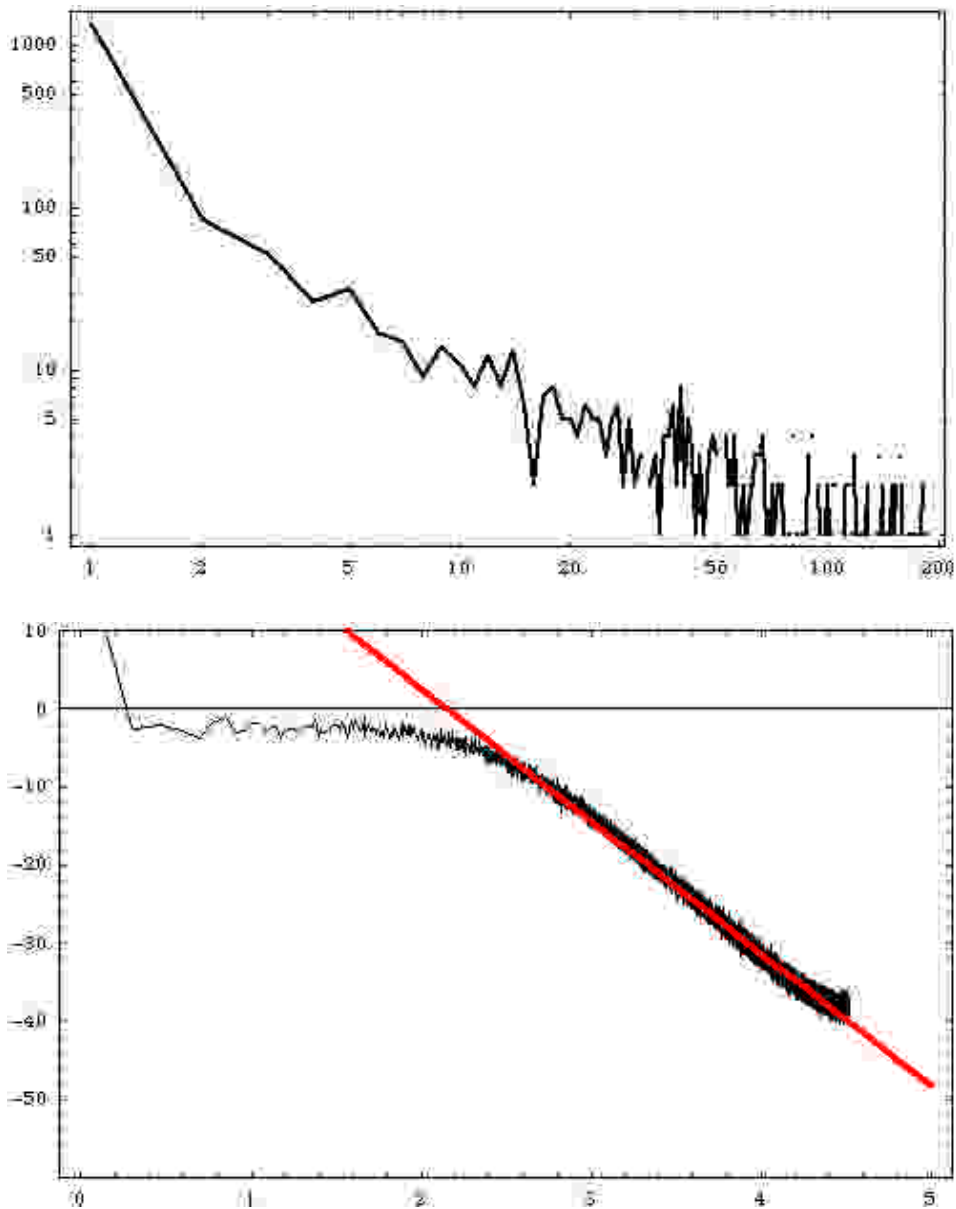

Figure 24: a. frequency distribution of relaxation rates; b. spectral density of the dipole fluctuation (dipole projected along a given direction) averaged over several molecules (dB(Debye, rms) vs. log(frequency (a.u.)). The distribution in a. closely follows a power-law, and therefore the results of section 2 apply to this case. The steep rise in b. at low frequency is due to an uncorrected DC component. The rise at high frequency is due to unfiltered high-frequencies (it is an aliasing effect). The red line shows a $1/f^{1.7}$ spectral density.

## 10. Numerical simulation $1/f^{\alpha}$ noise

We have seen simple dynamical processes that generate non-Gaussian $1/f$ noise, but in most cases one in interested in simulating Gaussian $1/f$ noise: there is a simple and reasonably fast algorithm that generates colored Gaussian noise [70], which is based on the following simple considerations a. the Fourier coefficients of a time series $\{x_k\}_{k=0,N-1}$ are given by the discrete Fourier transform

$$F_m = \frac{1}{\sqrt{N}} \sum_{k=0}^{N-1} x_k e^{-\frac{2\pi i k m}{N}} = \frac{1}{\sqrt{N}} \sum_{k=0}^{N-1} x_k \cos\frac{2\pi k m}{N} - i\frac{1}{\sqrt{N}} \sum_{k=0}^{N-1} x_k \sin\frac{2\pi k m}{N}; \qquad (33)$$

b. we see from expression () that if the variates $x_k$ are Gaussian, then the real and the imaginary part of $F_m$ are Gaussian as well; c. the periodogram of the time series is defined by

$$S_m = \frac{|F_m|^2}{N};$$

d. if we draw two normally distributed numbers with zero mean and unit variance and multiply them by $\sqrt{NS_m/2}$, then we obtain a spectral density with the desired properties; e. we want to generate a real time series, and therefore we require $F_{N-m} = F_m^*$, so that we have to generate only half of the Fourier coefficients $F_m$; f. inverse-Fourier transform the sequence $F_m$ to obtain the sequence $x_k$. Notice that if $N$ is a power of 2 one may use the FFT algorithm and the method is quite fast: using a moderate interpolation and sacrificing the high frequency component (which is buried in white noise in real experimental data) it is possible to generate very long time series [71]. Figure 25 shows examples of time series generated with this algorithm

## 11. *1/f* noise literature (including online resources)

There is a vast literature on *1/f* noise and we have barely touched the surface in this review. I have neglected several interesting approaches such as that based on the log-normal distribution [72,73] and resistor noise as a percolative process [74-79]. Many references to older material can be found in previous reviews [6,14,19,21,22,80,81].
There are also many interesting online resources, in particular there is a site dedicated to *1/f* noise, mantained by W. Li [82].

## 12. Conclusions

I conclude with a word of caution, these noises are beautiful physical phenomena, but we should beware of pitfalls: I have already mentioned the missing square in paper [45] (the authors of that paper wrongly claimed they had found a *1/f* spectral density): strangely enough, this mistake has been repeated many times, for instance in the original paper of BTW [32] and later in [83] (erratum in [84]).

In this review we have studied several mechanisms that produce fluctuations with a *1/f^α* spectral density: do we have by now an "explanation" of the apparent universality of flicker noises? Do we understand *1/f* noise? My impression is that there is no real mistery behind *1/f* noise, that there is no real universality and that in most cases the observed *1/f* noises have been explained by beautiful and mostly *ad hoc* models.

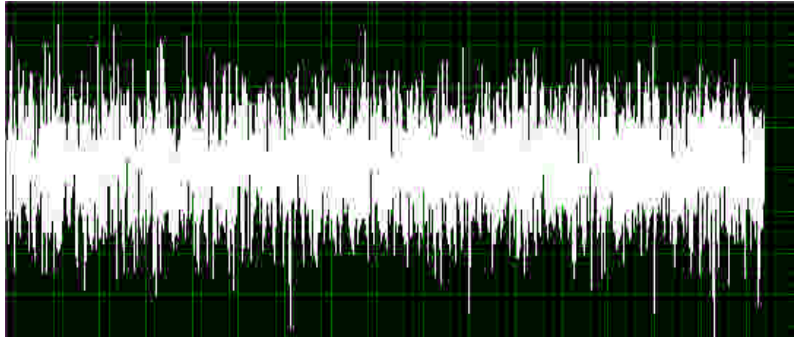
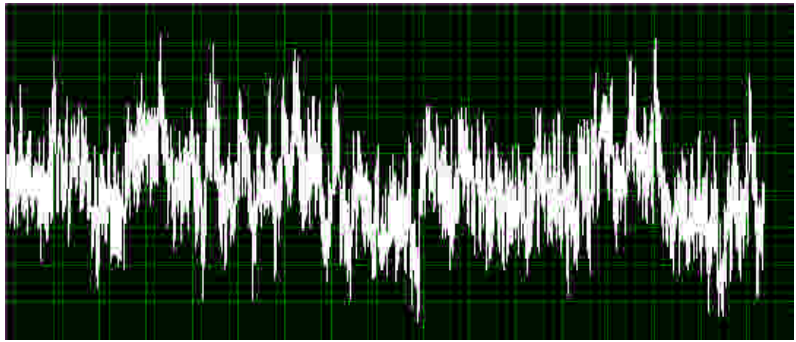
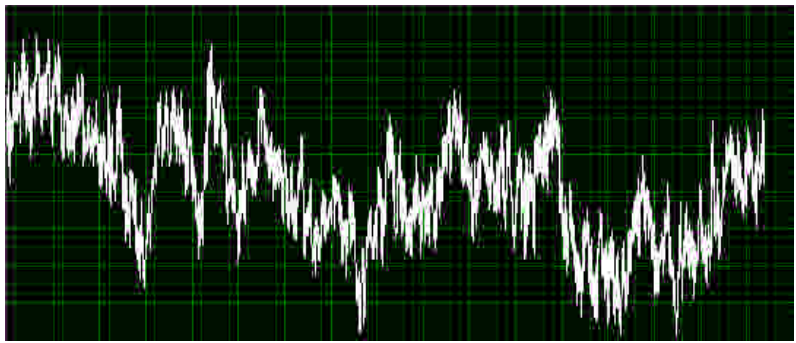
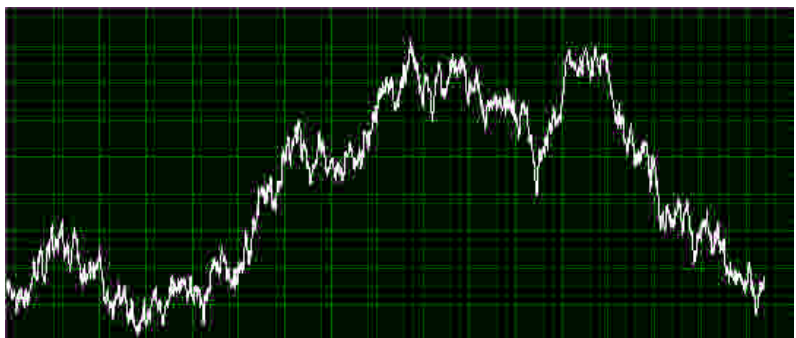

time

Figure 25: *1/f$^\alpha$* noise generated with the algorithm described in section 10, amplitude vs. time (both linear scales and arbitrary units); starting from top, $\alpha$ = 0, 1, 1.5 and 2.

# References

[1]   W. Schottky, Ann. der Phys. **57** (1918) 541; ibidem **68** (1922) 157.
[2]   C. A. Hartmann, Ann. der Phys. **65** (1921) 51.
[3]   J. B. Johnson, Phys. Rev. **26** (1925) 71.
[4]   W. Schottky, Phys. Rev. **28** (1926) 74.
[5]   B. Mandelbrot, *Fractals: Form, Chance and Dimension* (W. H. Freeman and Co. 1977).
[6]   W. H. Press, Comments Astrophys. **7** (1978) 103.
[7]   B. A. Taft et al., Deep Sea Research **21** (1974) 403.
[8]   C. Wunsch, Rev. Geophys. and Space Phys. **10** (1972) 1.
[9]   M. Gardner, Sci. Am. **238** (1978) 16.
[10]  R. F. Voss and J. Clarke, Nature **258** (1975) 317.
[11]  R.F. Voss and J. Clarke, J. Acoust. Soc. Am. **63** (1978) 258.
[12]  J. Bernamont, Ann. Phys. (Leipzig) **7** (1937) 71.
[13]  A. R. Butz, J. Stat. Phys. **4** (1972) 199.
[14]  A. van der Ziel, Adv. Electronics and Electron Phys. **49** (1979) 225.
[15]  B. Pellegrini, R. Saletti, P. Terreni and M. Prudenziati, Phys. Rev. **B27** (1983) 1233.
[16]  M. A. Caloyannides, J. Appl. Phys. **45** (1974) 307.
[17]  I. Flinn, Nature **219** (1968) 1356.
[18]  M. C. Wang and G. E. Uhlenbeck, Rev. Mod. Phys. **17** (1945) 323.
[19]  P. Dutta and P. M. Horn, Rev. Mod. Phys. **53** (1981) 497.
[20]  E. Milotti, Phys. Rev. **E51** (1995) 3087.
[21]  M. B. Weissman, Rev. Mod. Phys. **60** (1994) 657.
[22]  Sh. M. Kogan, Sov. Phys. Usp. **28** (1985) 170.
[23]  F. N. Hooge, Physica **60** (1972) 163.
[24]  F. N. Hooge, Phys. Lett. **A29** (1969) 139.
[25]  R. F. Voss and J. Clarke, Phys. Rev. **B13** (1976) 556.
[26]  H. G. E. Beck and W. P. Spruit, J. Appl. Phys. **49** (1978) 3384.
[27]  J. J. Brophy, J. Appl. Phys. **40** (1969) 567.
[28]  L. J. Greenstein and J. J. Brophy, J. Appl. Phys. **40** (1969) 682.
[29]  R. F. Voss, Phys. Rev. Lett. **40** (1978) 913.
[30]  P. J. Restle et al. Phys. Rev. **B31** (1985) 2254.
[31]  P. Bak, C. Tang and K. Wiesenfeld, Phys. Rev. Lett. **59** (1987) 381
[32]  P. Bak, C. Tang and K. Wiesenfeld, Phys. Rev. **A38** (1988) 364
[33]  S. Brown and G. Grüner, Sci. Am. April 1994, p.50.
[34]  R. E. Thorne, Phys. Today, May 1996, p.42.
[35]  G. Grüner, Rev. Mod. Phys. **60** (1988) 1129.
[36]  K. L. Schick and A. A. Verveen, Nature **251** (1974) 599.
[37]  B. Drossel and F. Schwabl, Phys. Rev. Lett. **69** (1992) 1629.
[38]  J. de Boer et al, Phys. Rev. Lett. **73** (1994) 906.
[39]  J. Davidsen and N. Lüthje, Phys. Rev. **E63** (2001) 063101-1.
[40]  D. Dhar, Physica **A263** (1999) 4.
[41]  K. P. O'Brien and M. B. Weissman, Phys. Rev. **A46** (1992) R4475.
[42]  H. M. Jaeger, C. Liu and S. R. Nagel, Phys. Rev. Lett. **62** (1989) 40.
[43]  S. R. Nagel, Rev. Mod. Phys. **64** (1992) 321.
[44]  H. J. Jensen, K. Christensen and H. C. Fogedby, Phys. Rev. **B40** (1989) 7425.
[45]  P. Helander et al, Phys. Rev. **E59** (1999) 6356.
[46]  F. Dalton and D. Corcoran, Phys. Rev. **E63** (2001) 061312-1.
[47]  N. Ferguson, Nature **408** (2000) 21.
[48]  B. Gutenberg and C. Richter, *Seismicity of the Earth and Associated Phenomena*, 2nd. ed. (Princeton Univ. Press, 1956).

[49] R. Burridge and L. Knopoff, Bull. Sism. Soc. Am. **57** (1967) 341.
[50] Northern California Earthquake Data Center, University of California,Berkeley Seismological Laboratory, http://quake.geo.berkeley.edu.
[51] B. Goss Levi, Phys. Today Nov. 1990, p. 17.
[52] J. M. Carlson and J. S. Langer, Phys. Rev. A40 (1989) 6470.
[53] J. M. Carlson, J. S. Langer and B. E. Shaw, Rev. Mod. Phys. **66** (1994) 657.
[54] http://simscience.org/crackling/Advanced/Earthquakes/EarthquakeSimulation.html.
[55] H. Kanamori and E. E. Brodsky, Phys. Today June 2001, p. 34.
[56] H. Kanamori and J. Mori, in *Problems in Geophysics for the New Millennium: A Collection of Papers in Honor of Adam M. Dziewonski,* E. Boschi, G. Ekström and A. Morelli eds. (Editrice Compositori, Bologna, 2000), p. 73.
[57] T. Geisel, A. Zacherl and G. Radons, Phys. Rev. Lett. **59** (1987) 2503.
[58] E. Lorentz, J. Atm. Sci. **20** (1963) 130.
[59] R. M. May, Nature **261** (1976) 459.
[60] P. Manneville and Y. Pomeau, Phys. Lett. **75A** (1979) 1.
[61] P. Manneville, Journal de Physique **41** (1980) 1235.
[62] J. E. Hirsch, B. A. Huberman and D. J. Scalapino, Phys. Rev. **A25** (1982) 519.
[63] M. J. Feingebaum, Physica **7D** (1983) 16.
[64] I. Procaccia and H. Schuster, Phys. Rev. **A28** (1983) 1210.
[65] A. Ben-Mizrachi et al. Phys. Rev. **A31** (1985) 1830.
[66] M. Peyrard, Phys. Rev. **E64** (2001) 011109-1.
[67] G. Careri and G. Consolini, Phys. Rev. **E62** (2000) 4454.
[68] G. Careri and E. Milotti, unpublished.
[69] D. H. Cobden et al., Phys. Rev. Lett. **69** (1992) 502.
[70] J. Timmer and M. Koenig, Astron. Astrophys. **300** (1995) 707.
[71] E. Milotti, "Algoritmi di simulazione di rumore colorato", presented at the 85th Congress of the Italian Physical Society, Pavia september 20-24 1999 (unpublished).
[72] B. West and M. Shlesinger, Am. Sci. **78** (1990) 40.
[73] B. West and M. Shlesinger, Int. J. Mod. Phys. **B3** (1989) 795.
[74] R. Rammal, C. Tannous and A.-M. S. Tremblay, Phys. Rev. **A31** (1985) 2662.
[75] R. Rammal and A.-M. S. Tremblay, Phys. Rev. Lett. **58** (1987) 415.
[76] L. M. Lust and J. Kakalios, Phys. Rev. **E50** (1994) 3431.
[77] L. M. Lust and J. Kakalios, Phys. Rev. Lett. **75** (1995) 2192.
[78] G. T. Seidler and S. A. Solin, Phys. Rev. **B53** (1996) 9753.
[79] G. T. Seidler, S. A. Solin and A. C. Marley, Phys. Rev. Lett. **76** (1996) 3049.
[80] V. Radeka, IEEE Trans. Nucl. Sci. **NS-16** (1969) 17.
[81] A. van der Ziel, Proc. IEEE **76** (1988) 233.
[82] W. Li, http://linkage.rockefeller.edu/wli/1fnoise/.
[83] H. J. S. Feder and J. Feder, Phys. Rev. Lett. 66 (1991) 2669.
[84] H. J. S. Feder and J. Feder, Phys. Rev. Lett. 67 (1991) 283.